\pdfoutput=1
\PassOptionsToPackage{bookmarks=false,colorlinks=true,linkcolor=blue,urlcolor=cyan,filecolor=green,citecolor=magenta,pdfstartview=FitV,hyperfootnotes=false,pdftitle={Analytic discrete self-similar solutions of Einstein--Klein--Gordon at large D},pdfauthor={Christian Ecker, Florian Ecker, Daniel Grumiller},pdfsubject={Analytic discrete self-similar solutions of Einstein--Klein--Gordon at large D},pdfkeywords={discrete self-similarity, spacetime crystals, critical collapse, large D expansion, general relativity, cosmic censorship, exact solutions, naked singularities},bookmarksopen=true}{hyperref}
\documentclass[aps,prl,10pt,twocolumn,superscriptaddress,amsmath,amssymb]{revtex4-2}

\usepackage{graphicx,bm,braket,mathrsfs,orcidlink,pgfplots}
\usepackage[normalem]{ulem}
\usepackage{tikz}
\usetikzlibrary{decorations.pathmorphing,backgrounds,calc,shapes,arrows,shapes.misc,shadows}
\tikzset{cross/.style={cross out, draw=black, minimum size=2*(#1-\pgflinewidth), inner sep=0pt, outer sep=0pt}, cross/.default={1pt}}
\usepgfplotslibrary{fillbetween}
\pgfplotsset{compat=1.18}

\newcommand{\eq}[2]{\begin{equation} #1 \label{#2} \end{equation}}

\DeclareMathOperator{\extdm}{d}
\newcommand{\extd}{\extdm \!}

\begin{document}


\preprint{TUW-25-03}

\def\mytitle{Analytic discrete self-similar solutions of Einstein--Klein--Gordon at large $\boldsymbol{D}$}
\title{\mytitle}

\author{\orcidlink{0000-0002-8669-4300}Christian Ecker}
\email{ecker@itp.uni-frankfurt.de}
\affiliation{%
Institute for Theoretical Physics, Goethe University\\ 60438, Frankfurt am Main, Germany
}%

\author{\orcidlink{0000-0002-0449-0081}Florian Ecker}
\email{fecker@hep.itp.tuwien.ac.at}
\affiliation{%
Institute for Theoretical Physics, TU Wien\\
Wiedner Hauptstrasse 8–10/136, A-1040 Vienna, Austria
}

\author{\orcidlink{0000-0001-7980-5394}Daniel Grumiller}
\email{grumil@hep.itp.tuwien.ac.at}
\affiliation{%
Institute for Theoretical Physics, TU Wien\\
Wiedner Hauptstrasse 8–10/136, A-1040 Vienna, Austria
}

\begin{abstract}
Discretely self-similar solutions govern critical gravitational collapse and have been known only numerically since Choptuik's pioneering work. We construct, in closed analytic form, an infinite family of such solutions of the Einstein-massless-Klein--Gordon system using the large-$D$ expansion. We characterize their structure and compare them with numerical critical solutions at finite $D$, identifying both universal features and distinctly large-$D$ behavior.
\end{abstract}

\maketitle


\section{Introduction}
\label{sec:1}

Black hole formation is a dramatic physical process that is hard to describe using only paper and pencil. Even imposing spherical symmetry and taking Einstein gravity with only a minimally coupled scalar field does not simplify the resulting PDEs sufficiently to permit analytic solutions in closed form \cite{Christodoulou:1986zr,Christodoulou:1986du,Christodoulou:1987vu,Goldwirth:1987nu}.

Choptuik constructed numerically the first solutions that can be arbitrarily close to the threshold of black hole formation \cite{Choptuik:1993jv} and discovered that they exhibit discrete self-similarity (DSS). The symmetry enhancement of the critical solution resembles the emergent conformal symmetry observed in physical systems at a second-order phase transition, see e.g.~\cite{Koike:1995jm,Gundlach:2002sx,Gundlach:2007gc} for comparisons between critical collapse and other physical systems. 

\begin{figure}
    \centering
    \includegraphics[width=\linewidth]{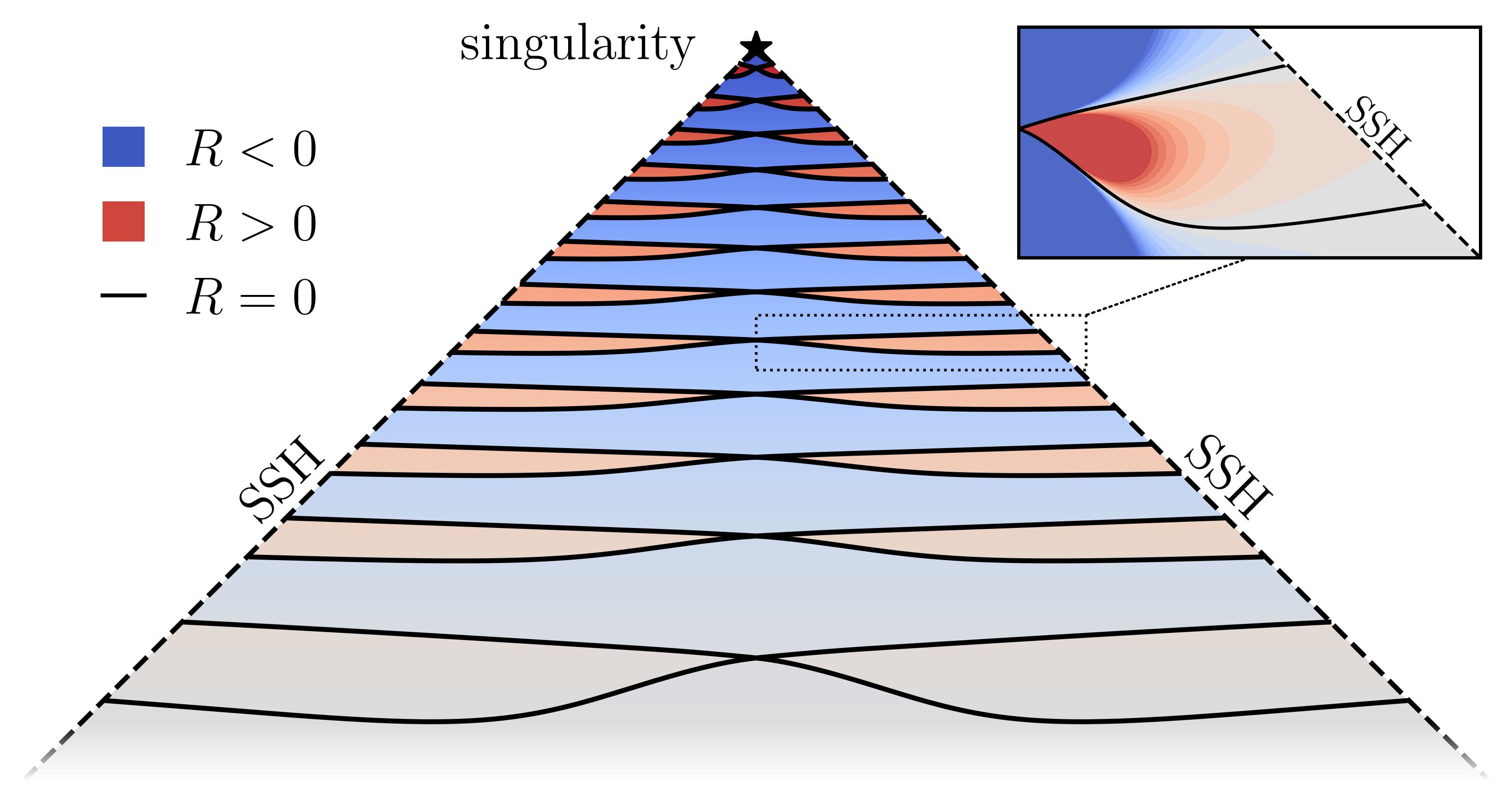}
    \caption{Illustration of the past patch of the Choptuik spacetime with DSS. The closer one moves towards the top, the more rapidly the Ricci scalar oscillates until the singularity is reached. The inset zooms into one of the fundamental domains to highlight the Ricci contours, i.e., the magnitude of the Ricci scalar is bigger in the center and smaller at the SSH. 
    }
    \label{fig:1}
\end{figure}

A related property is the existence of a self-similar horizon (SSH), i.e., a null hypersurface where the lattice vector $\partial_\tau$ associated with DSS, $\tau\to\tau+\Delta$ with echoing period $\Delta$, becomes null. The SSH is an ingoing null hypersurface and terminates in the center in a naked singularity. Emanating from the latter is an outgoing Cauchy horizon. See Fig.~\ref{fig:1} for a sketch of the past patch  of a critical solution with DSS. This means one can interpret the critical solution as (conformal to) a spacetime crystal where the lattice vector $\partial_\tau$ is timelike in the region between center and the SSH (called past region), spacelike between the SSH and the Cauchy horizon (called outer region), and lightlike on the SSH. 


This emergent discrete symmetry of the critical solution might, in principle, help to find it in closed form. Nevertheless, so far, the best one can do is to numerically construct the critical solution directly by imposing DSS as an input \cite{Gundlach:1995kd,Gundlach:1996eg,Martin-Garcia:2003xgm}. Since Einstein gravity has no dimensionless small parameter, there seems to be no obvious way to improve the situation.

However, if we allow deviations of the dimension $D$ from four to arbitrarily large values, we can use $1/D$ as a small dimensionless parameter \cite{Emparan:2013moa}. While it has been suggested in the past to exploit the large-$D$ expansion to understand critical collapse \cite{Emparan:2013xia,Rozali:2018yrv,Emparan:2020inr}, so far, the critical DSS solution seemed out of reach.

Numerically, going to large $D$ is problematic because radial gradients get steeper with increasing dimension \cite{Sorkin:2005vz,Bland:2005kk}. This numerical drawback has been turned into an advantage in the past to get analytical expressions for large-$D$ observables \cite{Emparan:2013moa,Emparan:2014cia,Emparan:2015gva,Emparan:2015hwa,Emparan:2015rva,Emparan:2020inr}. Our main goal is to show that the same is true for DSS solutions.

In the present work, we explicitly construct an infinite family of DSS solutions of spherically symmetric Einstein gravity in $D$ spacetime dimensions, minimally coupled to a massless scalar field $\psi$.

By construction, these solutions turn out to be regular between the center and the SSH and have a naked singularity in the center at the future endpoint of the SSH, just like for Choptuik's critical solution. We confront these solutions with numerical data to highlight the similarities and differences and show how to systematically improve features of the leading-order (LO) solutions by including next-to-leading-order (NLO) and next-to-next-to-leading-order (NNLO) corrections.

\section{Field equations}\label{sec:2}

Generalizing Choptuik's Eqs.~(3)-(5) in \cite{Choptuik:1993jv} to arbitrary spacetime dimensions $D>3$ yields
\begin{subequations}
    \label{eq:Chop3}
\begin{align}
   \frac{\partial_r a}{a} &= \frac{(D-3)\big(1-a^2\big)}{2r} + \frac{r}{2(D-2)}\,\big(\hat\Pi^2+\hat\Phi^2\big) \\
   \frac{\partial_r\alpha}{\alpha}  &=  \frac{\partial_r a}{a}+\frac{(D-3)\big(a^2-1\big)}{r}\\
    \partial_t\hat\Phi &= \partial_r\Big(\frac{\alpha}{a}\,\hat\Pi\Big)\qquad
    \partial_t\hat\Pi = \frac{1}{r^{D-2}}\,\partial_r\Big(r^{D-2}\,\frac{\alpha}{a}\,\hat\Phi\Big)
\end{align}
\end{subequations}
where the $D$-dimensional metric $\extd{s}^2=-\alpha^2(t,r)\extd{t}^2+a^2(t,r)\extd{r}^2+r^2\extd^2\Omega_{S^{D-2}}$ is spherically symmetric, i.e., $\extd^2\Omega_{S^{D-2}}$ is the metric of the round unit $S^{D-2}$. The quantities $\hat\Phi =\partial_r\psi$, $\hat\Pi = \frac{a}{\alpha}\,\partial_t\psi$ are suitably chosen first derivatives of the scalar field $\psi$. (We use units $c=8\pi G=1$, which differs from Choptuik's choice $c=G=1$.)

The coordinate system above is inconvenient for our purposes. Using the coordinate transformation \cite{Martin-Garcia:2003xgm} $\tau=-\ln(-t)$, $x=-r/t$ together with the redefinitions \footnote{%
The idea to rescale a variable like $a^2-1$ by a linear factor in the dimension in a large-$D$ context is due to Rozali and Way \cite{Rozali:2018yrv}. However, they also rescaled one of the coordinates, which makes the SSH inaccessible. Therefore, we do not follow their scaling regime but establish here a different one that allows a regular SSH in the large-$D$ expansion.} 
\begin{equation}
\Omega=\frac{(D-3)(a^2-1)}{x^2}\qquad\; f=\frac{\alpha}{a}\;\qquad \epsilon=\frac{1}{D-1}
\label{eq:Chop2}
\end{equation}
the field equations above read 
\begin{subequations}
    \label{eq:eom}
\begin{align}
x\partial_x\Omega + x^2\Omega^2 &= \frac1\epsilon\,\big((1-2\epsilon)\,\Pi^2-\Omega\big)+ \label{eq:eom1} \\
&\quad\;\Omega\,\big(\Pi^2x^2+x^4\,\Psi^2\big)+\Big(\frac1\epsilon-2\Big)\,x^2\,\Psi^2 \nonumber\\
   \frac1x\,\partial_x\ln f &=\Omega \label{eq:eom2} \\
   \big(\partial_\tau + x\,\partial_x\big)\,\Pi &= f\,x\,\partial_x \Psi + f\Big(\frac1\epsilon+x^2\,\Omega\Big)\,\Psi-\Pi \label{eq:eom3}\\
   \big(\partial_\tau + x\,\partial_x\big)\,\Psi &= f\,\frac1x\,\partial_x \Pi + f\,\Omega\, \Pi - 2 \Psi \label{eq:eom4}
\end{align}
\end{subequations}
where we introduced the first-order matter variables
\begin{equation}
\Pi=\sqrt{\frac{\epsilon}{1-\epsilon}}\frac1f\big(\partial_\tau+x\partial_x\big)\psi\qquad\;\Psi=\sqrt{\frac{\epsilon}{1-\epsilon}}\frac1x\partial_x\psi.
\label{eq:pipsi}
\end{equation}

The DSS property means that the four free functions that characterize our solutions, $\Omega, f, \Pi, \Psi$, are periodic in $\tau$ with some (echoing) period $\Delta$. The $D$-dimensional metric in the gauge above is given by
\begin{multline}
\extd s^2 = e^{-2\tau}\Big(1+\frac{\epsilon\,x^2\,\Omega}{1-2\epsilon}\Big)\big((x^2-f^2)\extd\tau^2-2x\extd\tau\extd x+\extd x^2\big)\\
+e^{-2\tau}\,x^2\,\extd\Omega^2_{S^{D-2}}\,.
\label{eq:metric}
\end{multline}
yielding a Ricci scalar $R=-(\frac{1}{\epsilon}+1)e^{2\tau}(\Pi^2-x^2\Psi^2)a^{-2}$ that grows like $\frac1\epsilon$ for large dimensions.

Assuming the SSH conditions $f(\tau,x=1)=1$, $f\geq x\geq0$, any solution to \eqref{eq:eom} must obey the inequalities
\begin{equation}
0\leq\Omega<\infty\qquad\qquad 0\leq x \leq f\leq 1\,.
\label{eq:ineqs}
\end{equation}

Our goal is to solve the equations \eqref{eq:eom} in an expansion of the dimension parameter $0<\epsilon\ll1$.

\section{Leading order solution at large $\boldsymbol{D}$}\label{sec:3}

We expand all fields as Taylor series in $\epsilon$, e.g., $\Omega=\Omega_{\textrm{\tiny LO}}+\epsilon\Omega_{\textrm{\tiny{NLO}}}+\epsilon^2\Omega_{\textrm{\tiny{NNLO}}}+\mathcal{O}(\epsilon^3)$ and analogously for the SSH function $f$ and the matter function $\Pi$. By contrast, as explained below, the other matter function $\Psi$ starts at one order higher, $\Psi=\epsilon\Psi_{\textrm{\tiny{LO}}}+\epsilon^2\Psi_{\textrm{\tiny{NLO}}}+\epsilon^3\Psi_{\textrm{\tiny{NNLO}}}+\mathcal{O}(\epsilon^4)$.

All coefficients of $\Omega$ are determined algebraically from \eqref{eq:eom1} in terms of $\Pi$. This is so because, to LO, the terms proportional to $1/\epsilon$ in the first line on the right-hand side of \eqref{eq:eom1} must cancel, i.e.,
\begin{equation}
\Omega_{\textrm{\tiny LO}} = \Pi^2_{\textrm{\tiny LO}}\,.
\label{eq:LO1}
\end{equation}
This mechanism persists for all subleading orders.

Moreover, $\Psi$ is determined without integration from $\Pi$ and $f$ using \eqref{eq:eom3}. Similarly to before, the reason for this is the $1/\epsilon$ coefficient on the right-hand side, yielding
\begin{equation}
\Psi_{\textrm{\tiny LO}} = \frac{D_1\,\Pi_{\textrm{\tiny LO}}}{f_{\textrm{\tiny LO}}}\qquad\qquad D_n:=\partial_\tau+x\partial_x+n\,.
\label{eq:LO2}
\end{equation}
Again, this mechanism persists for all subleading orders.

We only have two remaining PDEs that need to be solved. While in some general coordinate system these are still coupled, non-linear PDEs and hard to solve, in the gauge \eqref{eq:metric} they decouple at each order. 

Indeed, \eqref{eq:eom4} to LO simplifies to a non-linear first-order PDE for the matter function $\Pi_{\textrm{\tiny LO}}$,
\begin{equation}
\partial_x\Pi_{\textrm{\tiny LO}} = -x\,\Omega_{\textrm{\tiny LO}}\,\Pi_{\textrm{\tiny LO}} = -x\,\Pi^3_{\textrm{\tiny LO}} 
\label{eq:LO3}
\end{equation}
solved by
\begin{equation}
\Pi_{\textrm{\tiny LO}} = \frac{\beta(\tau)}{\sqrt{1+\beta^2(\tau)\,x^2}}
\label{eq:LO4}
\end{equation}
with an arbitrary integration function $\beta(\tau)$. We have adapted this integration function such that $\Pi_{\textrm{\tiny LO}}$ has no poles for any non-negative $x$. Also, we have chosen, without loss of generality, a branch of the square root function where $\Pi_{\textrm{\tiny LO}}$ is positive for positive $\beta$.

The final equation we need to solve is \eqref{eq:eom2}, which, to LO, reduces to
\begin{equation}
\partial_x \ln f_{\textrm{\tiny LO}} = x\,\Pi^2_{\textrm{\tiny LO}} = \frac{x\,\beta^2(\tau)}{1+\beta^2(\tau)\,x^2}\,.
\label{eq:LO5}
\end{equation}
We solve it with the boundary condition
\begin{equation}
\textrm{SSH:}\qquad f(\tau,x=1) = 1
\label{eq:LO6}
\end{equation}
that guarantees an SSH at $x=1$, see, e.g., the discussion in \cite{Martin-Garcia:2003xgm}. Thus, we have no further free integration function and obtain
\begin{equation}
f_{\textrm{\tiny LO}} = \sqrt{\frac{1+\beta^2(\tau)\,x^2}{1+\beta^2(\tau)}}\,.
\label{eq:LO7}
\end{equation}

\section{CSS and DSS solutions}

While $\beta(\tau)$ currently is unconstrained, if we choose it to be constant, we get continuously self-similar (CSS) solutions. They differ from the CSS solutions of Clark and Pimentel \cite{Clark:2025tqi}, see the Supplemental Material (SM). 

For periodic $\beta(\tau+\Delta)=\beta(\tau)$, we get DSS solutions. A prototypical choice is a function with zero average, a single zero in the interval $\tau\in[0,\Delta/2)$, and $\beta(\tau+\Delta/2)=-\beta(\tau)$. In this work we focus on such functions. 

Thus, we have achieved our main goal, namely, to find explicit (and remarkably simple) expressions for DSS solutions of the Einstein-massless--Klein--Gordon model, to LO at large $D$, Eqs.~\eqref{eq:LO1}-\eqref{eq:LO7}. We address now some of the properties and shortcomings of these solutions.

\section{Properties of leading order solution}\label{sec:4}

The solutions \eqref{eq:LO1}-\eqref{eq:LO7} are regular in the whole past region of Fig.~\ref{fig:1} and can also be extended into the outer region, until $x\to\infty$. To go beyond, one would need to change to a different coordinate system that allows one to reach the Cauchy horizon. This issue was resolved in \cite{Martin-Garcia:2003xgm}, and the same resolution works here as well, so we focus in the remainder of our work on the past region where we can use the convenient coordinate system \eqref{eq:metric} for any $\tau\in\mathbb{R}$ and any $x\in[0,1]$.

Another characteristic property of DSS solutions are the NEC lines \cite{Ecker:2024haw}, which are loci where the null energy condition saturates. The NEC lines coincide with lines of vanishing Ricci scalar and therefore also encode local geometry properties of the DSS solution. They are given by solutions to 
\begin{equation}
\textrm{NEC-lines:}\quad \Pi = \pm x\,\Psi
\label{eq:NEC} 
\end{equation}
and for our solutions \eqref{eq:LO1}-\eqref{eq:LO7} they emanate from NEC vertices located at single zeros of the function $\beta(\tau)$. 

A drawback of the $\mathcal{O}(1)$ approximation to \eqref{eq:NEC} is that the NEC lines remain straight horizontal lines in the chosen coordinates until they reach the SSH, without any NEC angle. If we ``cheat'' and insert $\mathcal{O}(1)$ on the left-hand side and the LO $\mathcal{O}(\epsilon)$ on the right-hand side of \eqref{eq:NEC}, we get the correct NEC angle, but the NEC lines remain straight lines. By contrast, at any finite $D$ the NEC lines curve downwards when emanating from the NEC vertices, see Fig.~3 in \cite{Ecker:2024haw}. Another aspect quite different from finite $D$ is that there is no unique DSS solution --- any periodic function $\beta(\tau)$ with any echoing period $\Delta$ solves our equations of motion. Finally, another drawback of the LO approximation is that the maxima of the function $f_{\textrm{\tiny LO}}$, located at zeros of $\beta$, remain at $f=1$ for all values of $x$. By contrast, at any finite $D$ these maxima have a characteristic $x$-profile \cite{Ecker:prep}. 

Below, we show that adding NLO and NNLO corrections addresses all the issues above.

\section{NLO solution}\label{sec:5}

We obtain $\Omega_{\textrm{\tiny NLO}}$ from \eqref{eq:eom1} 
and $\Psi_{\textrm{\tiny NLO}}$ from \eqref{eq:eom3},
\begin{subequations}
    \label{eq:angelinajolie}
\begin{align}
\Omega_{\textrm{\tiny NLO}} &= 2\Pi_{\textrm{\tiny LO}}\Pi_{\textrm{\tiny NLO}} - \omega_{\textrm{\tiny LO}}
\label{eq:NLO1} \\
\Psi_{\textrm{\tiny NLO}} &= \frac{D_1\Pi_{\textrm{\tiny NLO}}}{f_{\textrm{\tiny LO}}} - \big(x\partial_x+x^2\Pi^2_{\textrm{\tiny LO}} + \frac{f_{\textrm{\tiny NLO}}}{f_{\textrm{\tiny LO}}} \big)\Psi_{\textrm{\tiny LO}}\,.  
\label{eq:NLO2}
\end{align}
\end{subequations}
with $\omega_{\textrm{\tiny LO}}:=2\Pi^2_{\textrm{\tiny LO}}-2x^2\Pi^4_{\textrm{\tiny LO}}$. The two remaining PDEs
\begin{subequations}
\label{eq:PDEs}
\begin{align}
\Big(\frac1x\,\partial_x+3\Pi^2_{\textrm{\tiny LO}}\Big)\Pi_{\textrm{\tiny NLO}} &= \frac{D_2\Psi_{\textrm{\tiny LO}}}{f_{\textrm{\tiny LO}}} +\Pi_{\textrm{\tiny LO}}\,\omega_{\textrm{\tiny LO}} \\
\Big(\frac1x\,\partial_x -\Pi^2_{\textrm{\tiny LO}}\Big)\,f_{\textrm{\tiny NLO}} &= 2f_{\textrm{\tiny LO}}\,\Pi_{\textrm{\tiny LO}}\,\Pi_{\textrm{\tiny NLO}} - f_{\textrm{\tiny LO}}\,\omega_{\textrm{\tiny LO}}
\end{align}
\end{subequations}
are solved by (prime denotes $\partial_\tau$) 
\begin{align}
\Pi_{\textrm{\tiny NLO}} &= \frac{4\beta^2x^2(1+\beta^2)(\beta+\beta')^2+p_1\ln(1+\beta^2x^2)}{2\beta^3\,(1+\beta^2x^2)^{5/2}}  \\
f_{\textrm{\tiny NLO}} &= \frac{p_2+p_3\,\ln(1+\beta^2)+p_4\,\ln(1+\beta^2x^2)}{2\beta^4\,(1+\beta^2)^{3/2}\,(1+\beta^2x^2)^{3/2}}
    \label{eq:NLO7}
\end{align}
where $p_n$ are polynomials in $x^2$, $p_1=(1+\beta^2x^2)(\beta^3(\beta''-4\beta')-\beta^2(3\beta^{\prime2}+2)+\beta(\beta''-5\beta')-4\beta^{\prime2})$, $p_2=-(1-x^2)\beta^2[p_1+4\beta^3\beta'+2\beta^2\beta^{\prime2}+2\beta^2x^2(\beta^4+4\beta^3\beta^\prime+\beta^2(1+2\beta^{\prime2})+2\beta\beta^\prime+\beta^{\prime2})]$, $p_3=p_1(1+\beta^2x^2)$, $p_4=-p_1(1+\beta^2)$.

Remarkably, both integration functions are fixed, the one for $f_{\textrm{\tiny NLO}}$ by demanding the SSH condition \eqref{eq:LO6} and the one for $\Pi_{\textrm{\tiny NLO}}$ by redefining the LO integration function $\beta(\tau)$ such that $\Pi=\Pi_{\textrm{\tiny LO}}+\epsilon\Pi_{\textrm{\tiny NLO}}=\beta(\tau)$ at $x=0$.

\section{Echoing period at NLO}\label{sec:6}

Since the exact solution and the LO solution obey the convexity conditions \eqref{eq:ineqs}, we demand that also the NLO solution \eqref{eq:NLO1}-\eqref{eq:NLO7} is consistent with them. Interestingly, this is not the case automatically. As we show now, resolving this issue can determine the echoing period $\Delta$.

Consider an expansion of $f$ near the SSH and additionally expand close to a zero of $\beta$, 
\begin{equation}
f_{\textrm{\tiny LO}} + \epsilon\,f_{\textrm{\tiny NLO}} = 1-\epsilon\,(1-x)\,\beta\,\big(\beta''+3\beta'\big) + \dots\,.
\label{eq:fSSH}
\end{equation}
If $\beta''+3\beta'$ does not vanish simultaneously with $\beta$, the subleading term in \eqref{eq:fSSH} is positive on one side of the zero of $\beta$ and hence $f>1$, thereby contradicting one of the convexity conditions \eqref{eq:ineqs}. 

Thus, at NLO, not all functions $\beta(\tau)$ are allowed anymore. However, all functions $\beta(\tau/\Delta)$ with period $\Delta$ can be converted into admissible functions by adjusting
\begin{equation}
\textrm{echoing\;period:}\quad \Delta = \frac{|\beta''|}{3|\beta'|}\Big|_{\beta=0}\,.
\label{eq:Delta}
\end{equation}
The condition \eqref{eq:Delta} guarantees $\partial_\tau^2\beta(\tau/\Delta)+3\partial_\tau\beta(\tau/\Delta)$ has a single zero at the zero of $\beta$ so that $f\leq1$ can be fulfilled. Moreover, it excludes simple functions like $\beta\propto\cos(2\pi\tau/\Delta)$ where the second (or first) derivative has zeros coincident with the zeros of $\beta$.

\section{Example Solution}\label{sec:7}

In what follows, we analyze a specific critical solution up to (N)NNLO using 
\begin{equation}
\beta(\tau)=\cos(2\pi\tau)+\frac{\sin(6\pi\tau)}{A}\qquad\qquad A\approx15.9476
\label{eq:example}
\end{equation}
where the echoing period is fixed to $\Delta=1$. 

The choice \eqref{eq:example} is motivated by simplicity: we normalized the amplitude of the leading Fourier mode to one, fixed its phase to a cosine mode, and allowed only one higher Fourier mode with orthogonal phase, a sine mode. Since we know from numerical solutions at finite $D$ that the frequencies of higher modes are odd integer multiples of the fundamental frequency \cite{Ecker:prep} we took thrice the fundamental frequency as argument of the sine mode. The amplitude $1/A$ of the second term is chosen to satisfy the condition \eqref{eq:Delta} with $\Delta=1$. 

For each choice of $\beta$ and each given order, there is a lower bound on the dimension $D$ where the solution is (barely) viable; for the example \eqref{eq:example} to NNLO this value is $D=52$. For $D=51$ there is a region near the SSH where $f>1$, contradicting one of the convexity conditions \eqref{eq:ineqs}, see the SM and \cite{Mathematica:largeD} for details.

\begin{figure}[htb]
\centering
\includegraphics[width=\linewidth]{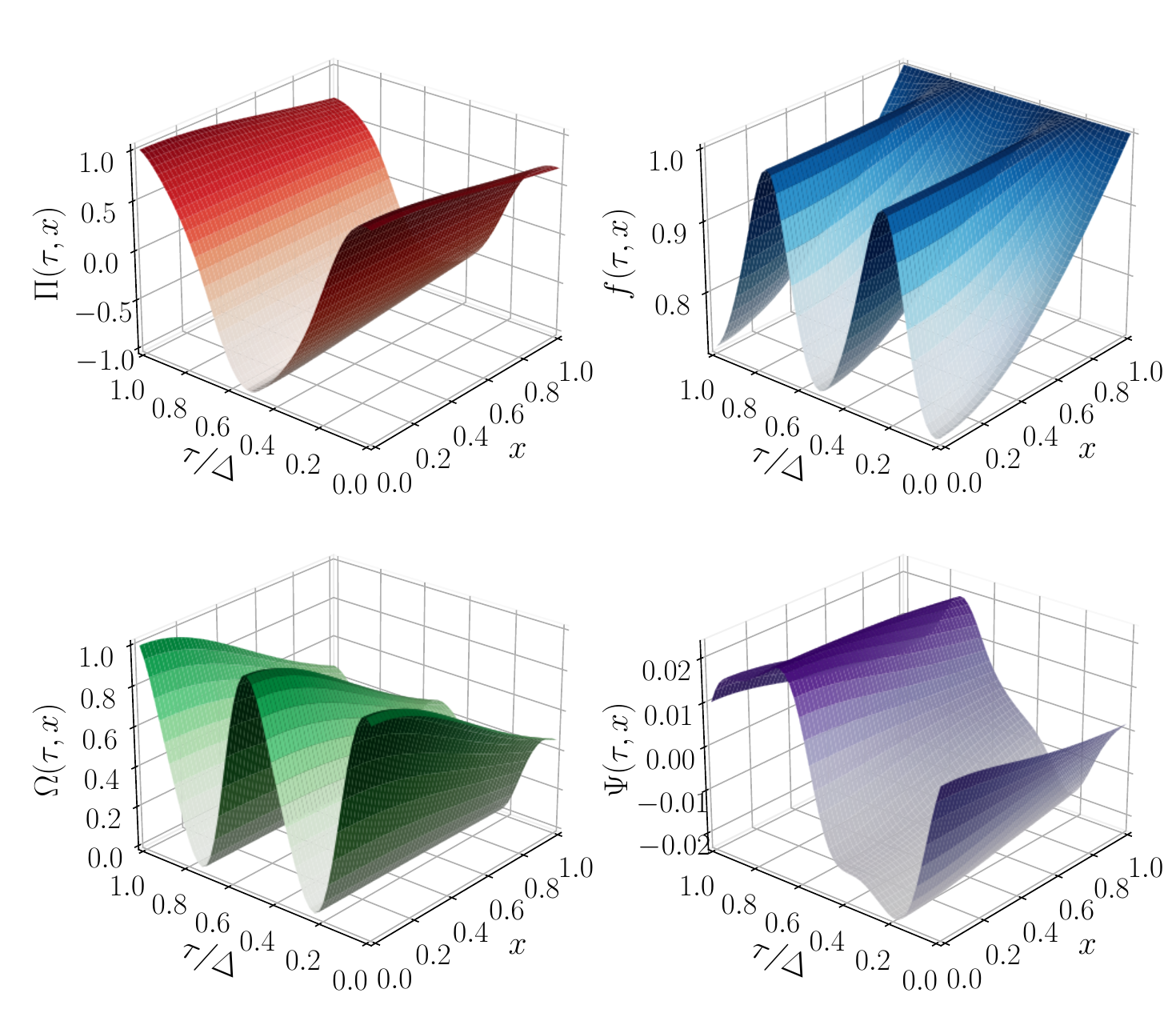}
\caption{NNLO solution for \eqref{eq:example} with $D=300$.}
\label{fig:fields}
\end{figure}

Figure~\ref{fig:fields} displays all fields of the NNLO solution evaluated at $D=300$ for the example \eqref{eq:example}. These plots can be compared and contrasted with the $D=4$ plots in \cite{Martin-Garcia:2003xgm}: e.g., the shape of the function $f$ (right upper graph) resembles the corresponding shape in $D=4$, except that our value of the maximum decays only marginally from $1$ near the SSH to slightly below $1$ in the center, see \eqref{eq:fmax} below, whereas in $D=4$ it decays to about $0.456$. 
%
%

\begin{figure}[htb]
\centering
\includegraphics[width=0.8\linewidth]{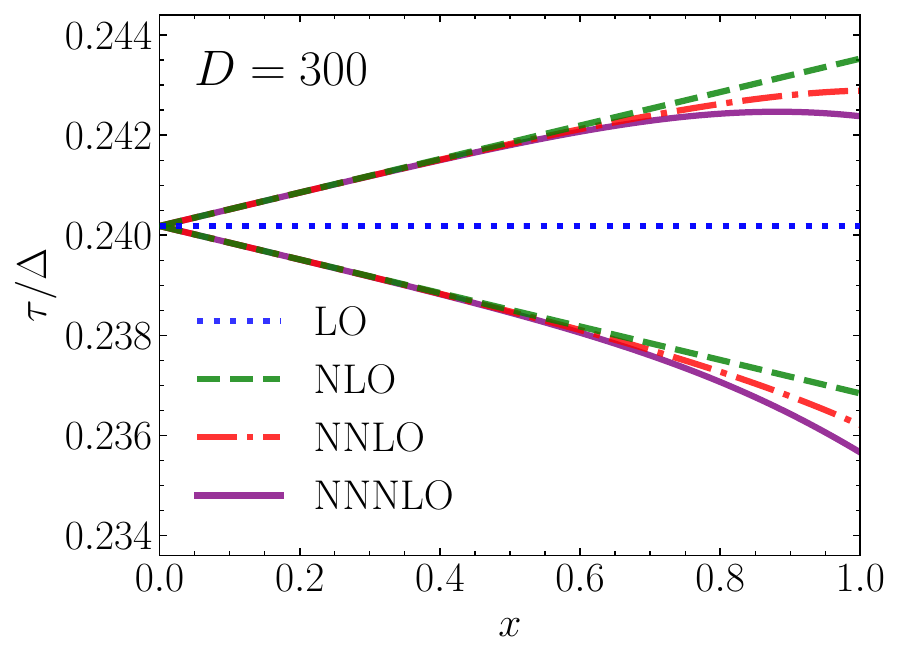}
\caption{NEC lines for \eqref{eq:example} with $D=300$.}
\label{fig:2}
\end{figure}

Figure~\ref{fig:2} shows the NEC lines given by solutions to \eqref{eq:NEC} for LO up to NNNLO for the same example. 
In the center, $x=0$, all curves are close to each other but near the SSH there are differences between these curves. This suggests that the large-$D$ expansion converges fast in the center and slowly at the SSH. The downward bending of the NEC lines observed in numerical simulations \cite{Ecker:2024haw,Ecker:prep} happens not before NNLO.
%
%

Finally, to LO and NLO the maxima of $f$ are at $f=1$ for all values of $x$. To NNLO we get
\begin{equation}
\textrm{max}\,f(\tau,\,x) = 1 - \frac{\epsilon^2}{4}\,\big(1-x^4\big)\,\big(\beta^{\prime}(\tau_0)\big)^2 + \mathcal{O}(\epsilon^3)
\label{eq:fmax}
\end{equation}
where $\tau_0$ is the NEC vertex time. The drop of the maximum of $f$ from $1$ to something proportional to $1/(D-1)^2$ is in qualitative agreement with numerical simulations at moderate values of $D$ \cite{Ecker:prep}.


\section{Summary and outlook}\label{sec:8}
We have provided a family of exact DSS solutions \eqref{eq:LO1}–\eqref{eq:LO7} of Einstein gravity in the large-$D$ expansion and showed how to systematically improve the solution order by order in $1/D$. To all orders, the input is a single free function of time $\beta(\tau)\sim\beta(\tau+\Delta)$ with arbitrary echoing period $\Delta$. We demonstrated that NLO corrections impose a consistency constraint \eqref{eq:Delta} on $\Delta$. For the example \eqref{eq:example}, we showed that several features of numerically obtained finite-$D$ solutions are captured once NNLO corrections are included, in particular the decay of the maximum of the SSH function $f$ from $1$ at the SSH to a smaller central value \eqref{eq:fmax} [see also Fig.~\ref{fig:fields}], as well as the NEC lines in Fig.~\ref{fig:2}.

Many rewarding questions remain open. Besides extending the perturbation theory past NNNLO (see the SM), it would be important to understand whether the $1/D$ series converges and, if so, with which radius of convergence. The NEC lines in Fig.~\ref{fig:2} to NNNLO suggest fast convergence near the center but slow convergence near the SSH. It is also plausible that higher orders impose further consistency relations analogous to \eqref{eq:Delta}; if infinitely many exist, they would strongly constrain $\beta(\tau)$ and $\Delta$, consistent with the numerical expectation that DSS solutions are unique at finite $D$ \cite{Gundlach:1998wm,Reiterer:2012hnr}. A better quantitative match to numerical data at moderate $D$ could be achieved by using more elaborate choices of $\beta(\tau)$ and by increasing the perturbative order.

Our discussion restricted to the past patch (Fig.~\ref{fig:1}). As in \cite{Martin-Garcia:2003xgm}, one may extend the solution through the outer patch to the future Cauchy horizon, and even study possible extensions beyond it. 

Moreover, we focused on solutions with a regular SSH. If we drop that assumption we find another branch of solutions, to LO given by
$f_{\textrm{\tiny{LO}}}=f(\tau)\sqrt{\beta^2(\tau)x^2-1}$, $\Omega_{\textrm{\tiny{LO}}}=$~Eq.\eqref{eq:LO1},
$\Pi_{\textrm{\tiny{LO}}}=\beta(\tau)/\sqrt{\beta^2(\tau)x^2-1}$, $\Psi_{\textrm{\tiny{LO}}}=$~Eq.\eqref{eq:LO2}.
where we assume $\beta^2(\tau)x^2>1$, which excludes $\beta$ having zeros. These solutions resemble the supercritical solutions above the threshold of black hole formation and contain a second arbitrary function, $f(\tau)$, besides $\beta(\tau)$. A detailed analysis of this branch would be interesting.

On the numerical side, it remains unclear how the echoing period $\Delta$ scales with $D$. Existing data \cite{Sorkin:2005vz,Bland:2005kk,blandthesis} suggest $\Delta\to0$ as $D\to\infty$, but the precise rate matters for the structure of the perturbative expansion \cite{Ecker:prep}. Relatedly, semiclassical backreaction effects \cite{Tomasevic:2025clf,Tomasevic:2025kqy}, problematic at finite $D$ \cite{Frolov:1999an}, may be consistently studied in the two-dimensional dilaton gravity formulation underlying the large-$D$ limit \cite{Emparan:2013xia}.

Finally, we have not addressed the Choptuik exponent $\gamma$, which at finite $D$ follows from linear perturbations around the DSS solution. Numerical and theoretical arguments suggest $\lim_{D\to\infty}\gamma=\tfrac12$, hinting at a CSS limit \cite{Clark:2025tqi}. We agree that the large-$D$ expansion is a powerful tool and emphasize that matching finite-$D$ data requires insisting on DSS and expanding around the DSS solutions constructed here.


\acknowledgments

\paragraph{Acknowledgments.} We thank Craig Clark for meeting with us and explaining in detail their work \cite{Clark:2025tqi}, Roberto Emparan for discussions on the large $D$ limit of general relativity, Carsten Gundlach for sharing and explaining his vintage code allowing the numerical construction of the DSS solution in $D=4$, and Maciej Maliborski for insightful feedback on our NEC angle discussion. DG is grateful to Peter Aichelburg and the other participants (Sascha Husa, Christiane Lechner, Michael P\"urrer, Jonathan Thornburg, and others) of the critical collapse seminar in 1999 for kindling his interest in this subject. 
CE acknowledges support by the DFG through the CRC-TR 211 ``Strong-interaction matter under extreme conditions'' -- project number 315477589 -- TRR 211. This work was supported by the Austrian Science Fund (FWF) [Grants DOI: \href{https://www.fwf.ac.at/en/research-radar/10.55776/P33789}{10.55776/ P33789}, \href{https://www.fwf.ac.at/en/research-radar/10.55776/P36619}{10.55776/P36619}, \href{https://www.fwf.ac.at/en/research-radar/10.55776/PAT1871425}{10.55776/PAT1871425}].  


\bibliographystyle{fullsort}
\bibliography{review}


\clearpage
\onecolumngrid
\setcounter{equation}{0}
\renewcommand{\theequation}{S\arabic{equation}}

\section{Supplemental Material}

In the Supplemental Material, we give more details on CSS solutions. Then, we obtain the scalar field from the first-order matter fields to NNLO. Next, we present the NNLO solution at large $D$ in terms of the single free function $\beta(\tau)$ that was introduced already at LO in the main text. Finally, we use this solution to discuss aspects of NEC lines, maxima/minima of the SSH function $f$ and the convexity conditions \eqref{eq:ineqs}, and plot sample results for the choice of $\beta$ made in the main text example. 

\subsection{More on CSS solutions}

Redefining the time as $\tau=t\,\Delta$ and assuming $\Delta\to0$ we obtain the LO CSS solution at large $D$ given by ($f_0,\beta\in\mathbb{R}$)
\begin{align}
    f_{\textrm{\tiny LO}} &= f_0\,\sqrt{1+\beta^2x^2} &
    \Omega_{\textrm{\tiny LO}} &= \frac{\beta^2}{1+\beta^2x^2} &
    \Pi_{\textrm{\tiny LO}} &= \frac{\beta}{\sqrt{1+\beta^2x^2}}&
    \Psi_{\textrm{\tiny LO}} &= \frac{\beta}{f_0\,(1+\beta^2x^2)^2}\,.
    \label{eq:CSS}
\end{align}
Compared to DSS solutions we have a second integration constant, $f_0$, since we dropped the assumption of having an SSH at $x=1$, so there is no longer a reason to enforce $f=1$ at $x=1$. If we map this solution to the one in Eq.~(2.29) of \cite{Clark:2025tqi}, then we see this works only for Minkowski space, $\beta=0$: Inserting the ansatz $x=\varphi_1(\hat x)$ and $\tau=-\omega+\varphi_2(\hat x)$ into our main text metric \eqref{eq:metric} with $\epsilon=0$ and matching with Eq.~(2.29) of \cite{Clark:2025tqi} yields the conditions $\varphi_2=\frac12\ln[(\varphi_1^2-f^2)/(2l^2)]$ and $\varphi_2^\prime=\varphi_1\varphi_1^\prime/(\varphi_1^2-f^2)$ (and a third condition, irrelevant for our argument). Differentiating the first of these conditions with respect to $\hat x$ yields $\varphi_2^\prime=(1-f_0^2\beta^2)\varphi_1\varphi_1^\prime/(\varphi_1^2-f^2)$ and thus consistency demands $f_0\beta=0$. Since $f_0$ is not allowed to vanish (otherwise, the metric would degenerate) the only option is $\beta=0$. We attribute the difference of our respective CSS solutions to different boundary conditions: \cite{Clark:2025tqi} assumed flat space boundary conditions at some value of the advanced time, whereas our CSS solutions are never flat for non-zero $\beta$. 

The supercritical branch is obtained by analytic continuation, $\beta\to i\beta$ and $f_0\to if_0$, yielding $f_{\textrm{\tiny LO}}=f_0\,\sqrt{\beta^2x^2-1}$ and analogous changes of $1+\beta^2\,x^2\to\beta^2x^2-1$ in all other fields in \eqref{eq:CSS}. The singular point $\beta^2x^2=1$ is sometimes used as a proxy for the apparent horizon, see, e.g., \cite{Rozali:2018yrv}. 

\subsection{Scalar field to NNLO}

While all the matter information is contained in the first-order fields $\Pi$ and $\Psi$, for some purposes it can be useful to recombine them into the scalar field $\psi$. Since this is a bit subtle, we explain here how this works for periodic $\beta(\tau)$.

Integrating the right Eq.~\eqref{eq:pipsi} with respect to $x$ determines the scalar field, which to LO is given by
\eq{
\psi_{\textrm{\tiny LO}}=\frac{\psi_0(\tau)}{\sqrt{\epsilon}}-\sqrt{\epsilon}\,\chi_0(\tau,\,x)\qquad\qquad\chi_0(\tau,\,x):=\frac{\sqrt{1+\beta^2(\tau)}\,(\beta(\tau)+\beta^\prime(\tau))}{2\beta^2(\tau)\,(1+\beta^2(\tau)x^2)} 
}{eq:LO42}
where we fix $\extd\psi_0(\tau)/\extd\tau=\beta(\tau)/\sqrt{1+\beta^2(\tau)}$ such that the LO solution is compatible also with the left Eq.~\eqref{eq:pipsi}. This ODE is integrated with the condition that the resulting function $\psi_0(\tau)$ is periodic with zero average (this is possible since we assume periodic $\beta(\tau)$ with zero average), which uniquely determines this function.

The second term in the scalar field \eqref{eq:LO42} is subleading but determined by the LO solutions for $\Pi$ and $\Psi$. This is so, because the expansion for $\Psi$ starts at one order lower than the one of $\Pi$. This pattern continues to subleading orders, 
\eq{
\psi(\tau,\,x) = \frac{1}{\sqrt{\epsilon}}\,\sum_{n=0}^\infty\psi_n(\tau)\,\epsilon^n + \sqrt{\epsilon}\,\sum_{n=0}^\infty\chi_n(\tau,\,x)\,\epsilon^n
}{eq:scalarallorders}
where the coefficient functions $\psi_n(\tau)$ and $\chi_n(\tau,\,x)$ are determined as follows: The right Eq.~\eqref{eq:pipsi} yields all the $\chi_n(\tau,\,x)$,
\eq{
\epsilon\,\Psi_{\textrm{\tiny LO}}+\epsilon^2\,\Psi_{\textrm{\tiny NLO}}+\epsilon^3\,\Psi_{\textrm{\tiny NNLO}} + \mathcal{O}(\epsilon^4) = \Big(\epsilon+\frac{\epsilon^2}{2}+\frac{3\epsilon^3}{8}\Big)\,\frac1x\,\partial_x\,\big(\chi_0+\chi_1\,\epsilon+\chi_2\,\epsilon^2\big) + \mathcal{O}(\epsilon^4) 
}{eq:chin}
where the integration functions implicit in $\chi_n$ are absorbed into $\psi_{n+1}$, and the left Eq.~\eqref{eq:pipsi} yields all the $\psi_n(\tau)$,
\begin{multline}
\Pi_{\textrm{\tiny LO}}+\epsilon\,\Pi_{\textrm{\tiny NLO}}+\epsilon^2\,\Pi_{\textrm{\tiny NNLO}} + \mathcal{O}(\epsilon^3) = \Big(1+\frac{\epsilon}{2}+\frac{3\epsilon^2}{8}\Big)\,\Big(\frac{1}{f_{\textrm{\tiny LO}}}-\frac{f_{\textrm{\tiny NLO}}\,\epsilon}{f^2_{\textrm{\tiny LO}}}+\frac{(f^2_{\textrm{\tiny NLO}}-f_{\textrm{\tiny LO}}f_{\textrm{\tiny NNLO}})\,\epsilon^2}{f^3_{\textrm{\tiny LO}}}\Big)\,x^2\,\big(\epsilon\,\Psi_{\textrm{\tiny LO}}+\epsilon^2\,\Psi_{\textrm{\tiny NLO}}\big)\\
+\Big(1+\frac{\epsilon}{2}+\frac{3\epsilon^2}{8}\Big)\,\Big(\frac{1}{f_{\textrm{\tiny LO}}}-\frac{f_{\textrm{\tiny NLO}}\,\epsilon}{f^2_{\textrm{\tiny LO}}}+\frac{(f^2_{\textrm{\tiny NLO}}-f_{\textrm{\tiny LO}}f_{\textrm{\tiny NNLO}})\,\epsilon^2}{f^3_{\textrm{\tiny LO}}}\Big)\,\partial_\tau\,\big(\psi_0+(\psi_1+\chi_0)\,\epsilon+(\psi_2+\chi_1)\,\epsilon^2\big)+\mathcal{O}(\epsilon^3)\,.
\label{eq:psin}
\end{multline}

\subsection{NNLO solution at large $\boldsymbol{D}$}

The scheme explained in the main text yielding the NLO solution can be applied exactly in the same way to each further subleading order, so that the large $D$ expansion can be improved perturbatively up to any desired order in $\epsilon=1/(D-1)$. We show here how this works at NNLO and provide explicit results in terms of elementary functions.

First, we solve algebraically for $\Omega_{\textrm{\tiny NNLO}}$,
\eq{
\Omega_{\textrm{\tiny NNLO}} = 2\Pi_{\textrm{\tiny LO}}\Pi_{\textrm{\tiny NNLO}} - \omega_{\textrm{\tiny NLO}}\qquad\qquad 
\omega_{\textrm{\tiny NLO}} := 2\big(x\partial_x+2\big)\big(\Pi_{\textrm{\tiny NLO}}\Pi_{\textrm{\tiny LO}}\big) - \Pi_{\textrm{\tiny NLO}}^2-x^2\Psi_{\textrm{\tiny LO}}^2 + 3x^2\Pi_{\textrm{\tiny LO}}^2\omega_{\textrm{\tiny LO}}
}{eq:s1}
and for $\Psi_{\textrm{\tiny NNLO}}$ (with $\mathcal{D}\Psi:=(x\partial_x+x^2\Pi^2_{\textrm{\tiny LO}})\Psi$),
\eq{
\Psi_{\textrm{\tiny NNLO}} = \frac{(\partial_\tau+x\partial_x+1)\Pi_{\textrm{\tiny NNLO}}}{f_{\textrm{\tiny LO}}}
-f_{\textrm{\tiny NNLO}}\,\frac{\Psi_{\textrm{\tiny LO}}}{f_{\textrm{\tiny LO}}}-f_{\textrm{\tiny NLO}}\,\frac{\Psi_{\textrm{\tiny NLO}}}{f_{\textrm{\tiny LO}}} -\frac{f_{\textrm{\tiny NLO}}}{f_{\textrm{\tiny LO}}}\mathcal{D}\Psi_{\textrm{\tiny LO}} -\mathcal{D}\Psi_{\textrm{\tiny NLO}}-2x^2\Pi_{\textrm{\tiny LO}}\Pi_{\textrm{\tiny NLO}}\Psi_{\textrm{\tiny LO}}+x^2\Psi_{\textrm{\tiny LO}}\omega_{\textrm{\tiny LO}}\,.
}{eq:s2}
The quantities on the right-hand sides are either known LO and NLO functions or determined below.

The remaining two PDEs have precisely the same differential operators as the NLO counterparts,
\begin{align}
\Big(\frac1x\,\partial_x+3\Pi^2_{\textrm{\tiny LO}}\Big)\,\Pi_{\textrm{\tiny NNLO}} &= \frac{\big(\partial_\tau+x\partial_x+2\big)\Psi_{\textrm{\tiny NLO}}}{f_{\textrm{\tiny LO}}} - \frac{f_{\textrm{\tiny LNO}}\,\big(\partial_\tau+x\,\partial_x+2\big)\Psi_{\textrm{\tiny LO}}}{f_{\textrm{\tiny LO}}^2}- 2\Pi_{\textrm{\tiny LO}} \Pi_{\textrm{\tiny NLO}}^2   + \omega_{\textrm{\tiny NLO}} \Pi_{\textrm{\tiny LO}} + \omega_{\textrm{\tiny LO}} \Pi_{\textrm{\tiny NLO}} \label{eq:s9}\\
\Big(\frac1x\,\partial_x -\Pi^2_{\textrm{\tiny LO}}\Big)\,f_{\textrm{\tiny NNLO}} &= 2f_{\textrm{\tiny LO}} \Pi_{\textrm{\tiny LO}}\Pi_{\textrm{\tiny NNLO}} + 2f_{\textrm{\tiny NLO}} \Pi_{\textrm{\tiny LO}}\Pi_{\textrm{\tiny NLO}} - f_{\textrm{\tiny LO}} \omega_{\textrm{\tiny NLO}}- f_{\textrm{\tiny NLO}} \omega_{\textrm{\tiny LO}}\label{eq:s10}
\end{align}
and can be solved in closed form, first \eqref{eq:s9} and then \eqref{eq:s10}. Again, there are no new integration functions since $f_{\textrm{\tiny NNLO}}$ is fixed by the SSH condition and the integration function implicit in $\Pi_{\textrm{\tiny NNLO}}$ is fixed by redefining the LO integration function such that the function $\Pi=\Pi_{\textrm{\tiny LO}}+\epsilon\,\Pi_{\textrm{\tiny NLO}}+\epsilon^2\,\Pi_{\textrm{\tiny NNLO}}$ at $x=0$ is given by the redefined function $\beta(\tau)$.

\newcommand{\apfour}{a^q}
\newcommand{\apthreel}{a^{cf}}
\newcommand{\apthree}{a^c}
\newcommand{\aptwotwo}{a^{s2}}
\newcommand{\aptwoltwo}{a^{sf2}}
\newcommand{\aptwol}{a^{sf}}
\newcommand{\aptwo}{a^s}
\newcommand{\aplfour}{a^{f4}}
\newcommand{\aplthree}{a^{f3}}
\newcommand{\apltwo}{a^{f2}}
\newcommand{\aplinear}{a^f}
\newcommand{\aconst}{a^z}

In terms of the single free function $\beta(\tau)$, we obtain the somewhat lengthy expressions 
\begin{align}
\Pi_{\textrm{\tiny NNLO}} &= \frac{\ln^2(1+\beta^2x^2)\,S_1 + \ln(1+\beta^2x^2)\ln(1+\beta^2)\,S_2}{8\beta^7(1+\beta^2x^2)^{5/2}} + \frac{\ln(1+\beta^2x^2)\,S_3 + \ln(1+\beta^2)\,S_4+S_5}{12\beta^7(1+\beta^2)(1+\beta^2x^2)^{9/2}} 
\label{eq:s3}\\
f_{\textrm{\tiny NNLO}} &= \frac{\ln^2(1+\beta^2x^2)\,S_6 + \ln(1+\beta^2x^2)\ln(1+\beta^2)\,S_7+ \ln^2(1+\beta^2)\,S_8  + \ln(1+\beta^2x^2)\,S_9 + \ln(1+\beta^2)\,S_{10}+S_{11}}{8\beta^8(1+\beta^2)^{5/2}(1+\beta^2x^2)^{5/2}}  
\label{eq:s4}
\end{align}
where the sums $S_i$ are generically of the form 
\eq{
S_i=\beta''''\apfour_i+\beta'''\beta^\prime\apthreel_i+\beta'''\apthree_i +\beta^{\prime\prime\,2}\aptwotwo_i+\beta''\beta^{\prime\,2}\aptwoltwo_i+\beta''\beta^\prime\aptwol_i+\beta''\aptwo_i +\beta^{\prime\,4}\aplfour_i+\beta^{\prime\,3}\aplthree_i+\beta^{\prime\,2}\apltwo_i+\beta^\prime\aplinear_i+\aconst_i
}{eq:s5}
with summands $a_i^\bullet$ that depend polynomially on $\beta$ and on $x^2$ with integer coefficients (for some sums, some summands vanish). To give an idea how they look like, we display all summands associated with the sum $S_1$ and otherwise refer to our Mathematica notebook \cite{Mathematica:largeD} that contains these results:
{\footnotesize
\begin{align}
\apfour_1 &= \beta^3(1+\beta^2)^2(1+\beta^2x^2) & \apthreel_1 &= -\beta^2(1+\beta^2)(1+\beta^2x^2)(12\beta^2+19) \\
\apthree_1 &= -\beta^3(1+\beta^2)(1+\beta^2x^2)(8\beta^2+9) &
\aptwotwo_1 &= -\beta^2(1+\beta^2)(9x^2\beta^4+(13x^2+6)\beta^2+10) \\
\aptwoltwo_1 &= \beta\big(72x^2\beta^6+3(77x^2+18)\beta^4+(163x^2+189)\beta^2+139\big) &
\aptwol_1 &= \beta^2\big(72x^2\beta^6+(205x^2+48)\beta^4+(134x^2+151)\beta^2+104\big) \\
\aptwo_1 &= \beta^3\big(16x^2\beta^6+16(3x^2+1)\beta^4+4(8x^2+9)\beta^2+20\big) &
\aplfour_1 &= -3\big(20x^2\beta^6+(80x^2+11)\beta^4+8(8x^2+7)\beta^2+48\big) \\ 
\aplthree_1 &= -3\beta\big(32x^2\beta^6+8(15x^2+1)\beta^4+(93x^2+58)\beta^2+53\big) &
\apltwo_1 &= -\beta^2\big(48x^2\beta^6+208x^2\beta^4+2(87x^2+26)\beta^2+51\big) \\ 
\aplinear_1 &= -4\beta^3\big(12x^2\beta^4+14x^2\beta^2-1\big) &
\aconst_1 &= -4\beta^4\big(2x^2\beta^2-1\big) 
\end{align}
}

Despite the poles in $1/\beta$ in individual expressions in \eqref{eq:s3} and \eqref{eq:s4}, there are no such poles in $\Pi_{\textrm{\tiny NNLO}}$ or $f_{\textrm{\tiny NNLO}}$ so that both fields are regular at zeros of $\beta$, i.e., at the NEC vertices.

We have also obtained the NNNLO solution, but due to the sheer size of the functions, we do not display them here and instead refer to our Mathematica notebook \cite{Mathematica:largeD}. The main two changes as compared to NNLO are the appearance of cubic powers in logarithms, i.e., terms containing a factor $\ln^3(1+\beta^2x^2)$ or $\ln^3(1+\beta^2)$ (or corresponding mixed terms), and the appearance of up to six $\tau$-derivatives of $\beta$. This suggests that this pattern continues to arbitrary higher orders, i.e., the N$^n$LO functions $\Pi$ and $f$ contain up to the $n^{\textrm{\tiny th}}$ power of logarithms and up to $2n$ $\tau$-derivatives of $\beta$.

\subsection{NEC lines to NNLO and beyond}

We solve now the NEC conditions \eqref{eq:NEC} up to NNLO. To LO, they reduce to a straight horizontal line at the NEC vertex time $\tau_0$ given by $\beta(\tau_0)=0$. To NLO, we have instead
\eq{
\Pi_{\textrm{\tiny LO}}^{(1)} + \epsilon\,\Pi_{\textrm{\tiny NLO}}^{(0)} = \pm\epsilon\,x\,\Psi_{\textrm{\tiny LO}}^{(0)} \qquad\qquad \Pi_{\textrm{\tiny LO}}^{(1)} =\beta^\prime(\tau-\tau_0)\qquad \Pi_{\textrm{\tiny NLO}}^{(0)}=\frac{x^2}{2}(3\beta^\prime+\beta'')=0\qquad \Psi_{\textrm{\tiny LO}}^{(0)} = \beta^\prime
}{eq:SMNEC1}
where the superscripts denote the order in $\tau-\tau_0$ up to which we expand the corresponding term. All expressions with $\tau$-derivatives of $\beta$ are evaluated at the NEC vertex time $\tau_0$. We assume $\beta^\prime\neq 0$, which is typically true for any random function and definitely true in all numerical simulations of DSS solutions at finite $D>3$ \cite{Ecker:prep}. Solving the linear equation \eqref{eq:SMNEC1} for $\tau$ yields straight lines, $\tau=\tau_0\pm\epsilon\,x$, establishing the correct NEC angle: The two spacelike tangent vectors $n_\pm^\mu\partial_\mu=\pm\partial_\tau+\frac1\epsilon\,\partial_x$ of the NEC lines have a relative rapidity $\xi=2\epsilon$, and taking the Gudermanian thereof yields the LO NEC angle $\alpha=2\epsilon$ \cite{Ecker:2024haw}.

To NNLO, we get the NEC line conditions
\begin{subequations}
    \label{eq:NNLONEC}
\begin{align}
\Pi_{\textrm{\tiny LO}}^{(2)} + \epsilon\,\Pi_{\textrm{\tiny NLO}}^{(1)} + \epsilon^2\,\Pi_{\textrm{\tiny NNLO}}^{(0)} &= \pm\epsilon\,x\,\big(\Psi_{\textrm{\tiny LO}}^{(1)}+\epsilon\,\Psi_{\textrm{\tiny NLO}}^{(0)}\big) &  
\Pi_{\textrm{\tiny NLO}}^{(1)} &= \frac{x^2}{2}\,(\tau-\tau_0)\,\big(\beta'''+(1-2x^2)\beta^{\prime\,3}-7\beta'\big) 
\label{eq:SMNEC2}\\
\Pi_{\textrm{\tiny LO}}^{(2)} &= \beta^\prime\,(\tau-\tau_0)\,\big(1-\frac32(\tau-\tau_0)\big) & \Pi_{\textrm{\tiny NNLO}}^{(0)} &= \frac{x^4}{8}\,\big(\beta''''+10\beta'''-6\beta^{\prime\,3}-55\beta^\prime\big) 
\label{eq:SMNEC3} \\
\Psi_{\textrm{\tiny LO}}^{(1)} &= \beta^\prime\,\big(1-2(\tau-\tau_0)\big) &  \Psi_{\textrm{\tiny NLO}}^{(0)} &= \frac{x^2}{2}\,\big(\beta'''+(1-2x^2)\,\beta^{\prime\,3}-7\beta^\prime\big)\,.
    \label{eq:SMNEC4}
\end{align}
\end{subequations}
Solving the quadratic NEC-line equations \eqref{eq:NNLONEC} directly yields two branches of solutions. On one branch, the solution connects smoothly to the NLO NEC-lines for small $\epsilon$, $\tau=\tau_0 + \mathcal{O}(\epsilon)$, on the other it does not, $\tau=\tau_0+\frac23+\mathcal{O}(\epsilon)$. Alternatively and more simply, we insert the NLO result plus a quadratic correction, $\tau=\tau_0\pm\epsilon\,x+\epsilon^2\,\delta\tau_{\textrm{\tiny NNLO}}$, yielding a linear equation for $\delta\tau_{\textrm{\tiny NNLO}}$. In either version we obtain the NNLO NEC lines
\eq{
\textrm{NNLO\;NEC\;lines:}\qquad \tau=\tau_0 \pm \epsilon\,x - \frac{\epsilon^2\,x^2}{2} - \frac{\epsilon^2\,x^4}{8\beta^\prime}\,\big(\beta''''+10\beta'''-6\beta^{\prime\,3}-55\beta^\prime\big)\,.
}{eq:SMNEC5}
The first quadratic term in $\epsilon$ is independent from $\beta$ and yields the characteristic downward bending of NEC lines near the center observed in numerical simulations \cite{Ecker:prep}. Since both NEC lines receive the same quadratic corrections in $\epsilon$, the NEC angle receives no $\mathcal{O}(\epsilon^2)$ corrections, in accordance with the derivation in \cite{Ecker:2024haw}.

Is is straightforward (though increasingly lengthy) to calculate the NEC lines to higher order. Schematically, to order N$^n$LO with positive integer $n$, we need $\Pi_{\textrm{\tiny LO}}^{(n)}$, $\Pi_{\textrm{\tiny NLO}}^{(n-1)}$, $\dots$, $\Pi_{\textrm{\tiny N}^n\textrm{\tiny LO}}^{(0)}$, $\Psi_{\textrm{\tiny LO}}^{(n-1)}$, $\Psi_{\textrm{\tiny NLO}}^{(n-2)}$, $\dots$, $\Psi_{\textrm{\tiny N}^{n-1}\textrm{\tiny LO}}^{(0)}$ and therefore end up with NEC line equations of order $n$ in $\tau-\tau_0$. Out of the $n$ different solutions, we take the unique one that connects smoothly to the lower orders. Alternatively and more simply, we make the ansatz $\tau=\tau_0\pm\epsilon\,x+\epsilon^2\,\delta\tau_{\textrm{\tiny NNLO}}+\epsilon^3\,\delta\tau_{\textrm{\tiny NNNLO}}+\dots+\epsilon^n\,\delta\tau_{\textrm{\tiny N}^n\textrm{\tiny LO}}$ and solve a linear equation for $\delta\tau_{\textrm{\tiny N}^n\textrm{\tiny LO}}$, using all the lower order contributions as an input. 

We calculated the NEC lines to NNNLO and found the following result.
\begin{multline} \tau = \underbrace{\textcolor{green}{\tau_0} \textcolor{magenta}{\pm \epsilon\,x} \textcolor{orange}{- \frac{\epsilon^2\,x^2}{2} - \frac{\epsilon^2\,x^4}{8\beta^\prime}\,\big(\beta''''+10\beta'''-6\beta^{\prime\,3}-55\beta^\prime\big)}}_{\textrm{See\;Eq.~\eqref{eq:SMNEC5}}}\textcolor{black}{-\frac{\epsilon^3\,\Pi^{(0)}_{\textrm{\tiny NNNLO}}}{\beta^\prime}}\textcolor{red}{\pm\frac{\epsilon^3\,x}{4}\,\beta^{\prime\,2}}\textcolor{black}{
\pm\frac{\epsilon^3\,x^3}{6\beta^\prime}\,\big(-4\beta'''-3\beta^{\prime\,3}+30\beta^\prime\big)} 
\\\textcolor{black}{\pm\frac{\epsilon^3x^5}{4\beta^\prime}\big(2\beta''''+20\beta'''-3\beta^{\prime\,3}-110\beta^\prime\big) + \frac{\epsilon^3x^4}{16\beta^{\prime\,2}}\big(4\beta^\prime + x^2(\beta'''+\beta^{\prime\,3}-7\beta^\prime)+2x^4\beta^{\prime\,3}\big)\big(\beta''''+10\beta'''-6\beta^{\prime\,3}-55\beta^\prime\big)}\\
\textcolor{black}{\Pi^{(0)}_{\textrm{\tiny NNNLO}}=\frac{x^2\,\beta^{\prime\,2}}{48}\,\big(\beta''''+10\beta'''-6\beta^{\prime\,3}-79\beta^\prime\big)+\frac{x^4}{32}\,\big(3\beta''''\beta^{\prime\,2}+6\beta'''\beta^{\prime\,2}-8\beta''''-80\beta'''+6\beta^{\prime\,5}++99\beta^{\prime\,3}+440\beta^\prime\big)}\\
\textcolor{black}{+\frac{x^6}{48}\,\big(\beta^{(6)}+21\beta^{(5)}+25\beta''''\beta^{\prime\,2}+120\beta'''\beta^{\prime\,2}+175\beta''''+735\beta'''-30\beta^{\prime\,5}-465\beta^{\prime\,3}-3108\beta^\prime\big)}\\
\textcolor{black}{-\frac{33x^8\,\beta^{\prime\,2}}{64}\,\big(\beta''''+10\beta'''-6\beta^{\prime\,3}-55\beta^\prime\big)}
\label{eq:SMNEC6}
\end{multline}
We have color coded the terms into \textcolor{green}{LO}, \textcolor{magenta}{NLO}, \textcolor{orange}{NNLO}, \textcolor{black}{generic NNNLO}, and \textcolor{red}{NNNLO linear in $x$}. A key feature that was absent at lower orders is the appearance of a correction to the NEC angle from the term $\textcolor{red}{\pm\frac{\epsilon^3\,x}{4}\,\beta^{\prime\,2}}$, the only term linear in $x$ apart from the NLO term $\textcolor{magenta}{\pm\epsilon\,x}$. This is precisely the behavior expected from the analysis in general dimension, see the discussion section of \cite{Ecker:2024haw}. Apart from the $\mathcal{O}(\epsilon^3)$ change of the NEC angle, the new correction terms have properties expected from our large-$D$ perturbation theory, in particular, terms with fifth and sixth derivatives of $\beta$.

\newpage

We summarize distinctive features of the NEC lines at the first few orders:
\begin{itemize}
    \item \textbf{LO.} We obtain the correct NEC vertex time but nothing else. The NEC lines degenerate to a single horizontal line.
    \item \textbf{NLO.} We additionally obtain the correct LO NEC angle. The NEC lines are both straight lines in the chosen coordinates.
    \item \textbf{NNLO.} We additionally get the correct downward bending of the NEC lines near the center. The NEC angle and the distance between the NEC lines remain unchanged compared to the NLO result.
    \item \textbf{NNNLO.} We additionally get an $\mathcal{O}(\epsilon^3)$ correction to the NEC angle, in agreement with \cite{Ecker:2024haw}. Moreover, the distance between the NEC lines changes compared to the NLO result.
\end{itemize}

\begin{figure}[htb]
\centering
\includegraphics[width=0.32\linewidth]{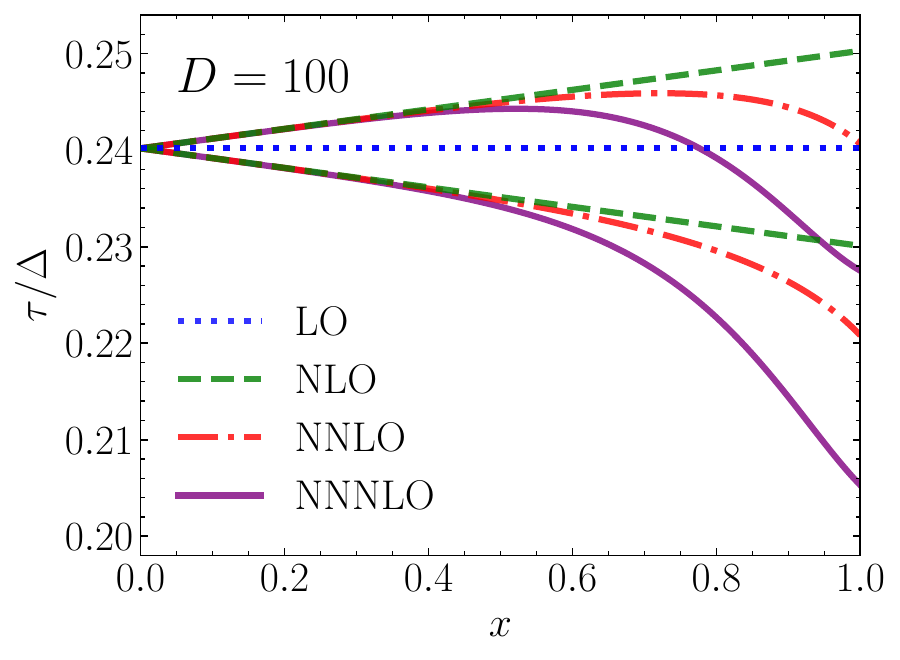}\quad
\includegraphics[width=0.32\linewidth]{necD300.pdf}\quad
\includegraphics[width=0.32\linewidth]{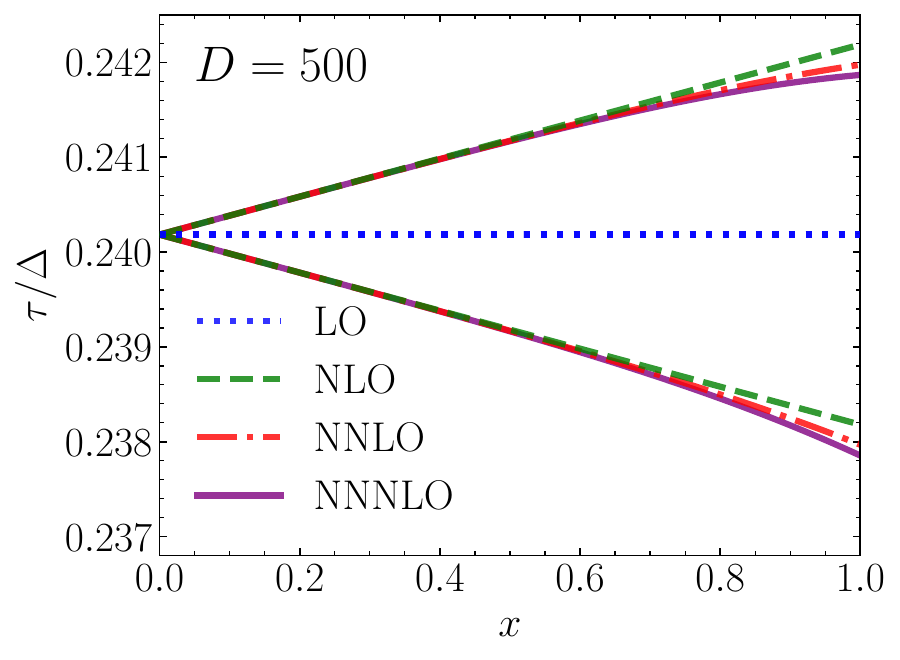}
\caption{NEC lines for \eqref{eq:example} with $D=100,300,500$.}
\label{fig:NECconvD}
\end{figure}

The NEC lines provide a useful probe of the convergence of the large-$D$ expansion at fixed spacetime dimension. To assess this, in Fig.~\ref{fig:NECconvD} we compare the NEC lines obtained at LO through NNNLO for three representative dimensions, ($D = 100, 300,500$), using the example solution \eqref{eq:example}. For the example shown, at $D=100$ (left panel) the NNNLO corrections exceed the NNLO ones, most clearly for $x \gtrsim 0.5$, with the largest deviation occurring at the SSH ($x=1$), indicating that the large-$D$ expansion is not convergent for $D \lesssim 100$. At $D \approx 300$ (middle panel), the NNLO and NNNLO corrections are comparable in magnitude, while at $D=500$ the NNNLO corrections are clearly smaller than the NNLO ones, indicating convergence of the series for this example at $D \gtrsim 500$ over the full range of $x$.

\subsection{Convexity conditions, maxima, and minima of SSH function $\boldsymbol{f}$ at NNLO}

We consider now the convexity conditions \eqref{eq:ineqs} to NNLO, starting with their proof to all orders.

The left two inequalities \eqref{eq:ineqs} follow from \eqref{eq:eom1}: At $x=0$, solving the quadratic equation for $\Omega$ and taking the positive root yields\eq{
\Omega(\tau,x=0)=\frac{1}{2\epsilon}\big(\sqrt{1+4\epsilon(1-2\epsilon)\,\Pi^2(\tau,x=0)}-1\big)\geq 0 
}{eq:lalapetz}
which is non-negative since $\epsilon<\frac12$ for $D>3$ and $\Pi$ is real. At positive $x$, the gradient of $\Omega$ is always non-negative as long $\Omega$ is small, since in this limit we get 
\eq{
x\partial_x\Omega=\frac1\epsilon\,(1-2\epsilon)\,(\Pi^2+x^2\,\Psi^2)+\mathcal{O}(\Omega)\geq 0\,. 
}{eq:whatever}
Therefore, $\Omega$ can never cross the line $\Omega=0$. Finally, if $\Omega$ became very large we would get $x\partial_x\Omega=-\Omega^2\,(1+\mathcal{O}(1/\Omega))<0$. Hence $\Omega$ would start decreasing and can never reach infinity. 

The right inequalities are mostly trivial: $0\leq{x}\leq1$ is just our coordinate domain and $x\leq{f}$ is part of our SSH conditions, making sure that there is no SSH for $x<1$. The only slightly non-trivial condition is $f\leq1$. It follows from the other SSH condition, $f(\tau,x=1)=1$, together with monotonicity implied by \eqref{eq:eom2}, $\partial_x{f}=x\,\Omega\geq0$.

To NLO, the convexity condition $f\leq1$ imposes a non-trivial constraint \eqref{eq:Delta}, which we have derived by expanding $f$ near the SSH \eqref{eq:fSSH}. We extend here this expansion to NNLO, focusing on the behavior of $f$ near its extrema, since they are most prone to violating $f\leq1$.

To LO, the extrema of $f_{\textrm{\tiny LO}}$ are located at zeros of 
\eq{
f^\prime_{\textrm{\tiny LO}} = \frac{(x^2-1)\,\beta\beta^\prime}{(1+\beta^2)^{3/2}\,\sqrt{1+\beta^2x^2}}
}{eq:s8}
i.e., if either $\beta$ or $\beta^\prime$ vanish. The extrema at $\beta=0$ are always maxima because the second derivative is negative there for $x<1$,
\eq{
f''_{\textrm{\tiny LO}}\big|_{\beta=0} = (x^2-1)\,\beta^{\prime\,2} \leq 0\,.
}{eq:s9a}
The other extrema can be either maxima or minima, depending on the sign of 
\eq{
f''_{\textrm{\tiny LO}}\big|_{\beta^\prime=0} = \frac{(x^2-1)\,\beta\beta''}{(1+\beta^2)^{3/2}\,\sqrt{1+\beta^2x^2}}
}{eq:s10a}
i.e., $\beta\beta''\leq 0$ ($\beta\beta''\geq 0$) leads to minima (maxima) at $\beta'=0$. At (N)NLO the relations above receive $\epsilon$-dependent modifications.

Close to the SSH these modification may lead to violations of the convexity condition $f\leq 1$ if the dimension is not large enough. We can quantify this at NLO by expanding near the SSH and close to an extremum with $\beta^\prime=0$,
\eq{
f=1-\frac{(1-x)\,\beta^2}{1+\beta^2}-\frac{2\epsilon\,(1-x)\,\beta^2}{(1+\beta^2)^2}-\frac{\epsilon\,(1-x)\,\ln(1+\beta^2)\,(\beta^2\beta''+\beta''-2\beta)}{\beta\,(1+\beta^2)^2} + \mathcal{O}(1-x)^2 + \mathcal{O}(\epsilon^2) + \mathcal{O}(\beta^\prime)^2 + \mathcal{O}(\epsilon\,\beta^\prime)\,.
}{eq:s9b}
The last displayed term can have either sign and thus puts an upper bound on $\epsilon$ to guarantee that $f$ never exceeds $1$. We can provide an order-of-magnitude estimate of the lowest dimension possible by requiring that $\epsilon\beta''|_{\beta^\prime=0}\sim\mathcal{O}(1)$, which is the scale at which the would-be-subleading term dispalyed above can compete with the $\mathcal{O}(1)$ terms if we assume $\beta\sim\mathcal{O}(1)$. For the main text example we get approximately $\beta''|_{\beta^\prime=0}\approx 4\pi^2$ and hence expect that the dimension has to be at least $40$. A more detailed analysis shows that for this example, including NNLO corrections, the minimal dimension is $52$, i.e., in the order of magnitude expected from this simple NLO estimate.

The maxima are fine at LO and NLO but at NNLO, we can again violate the convexity condition $f\leq 1$ if the dimension is not sufficiently large. We quantity this at NNLO by expanding near the SSH and around vanishing $\beta$,
\eq{
f = 1 - \frac{\epsilon^2\,(1-x)}{4}\,\big(4\beta^{\prime\,2}+\beta(\beta''''+10\beta'''-6\beta^{\prime\,3}-47\beta^\prime)\big)
+ \mathcal{O}(1-x)^2 + \mathcal{O}(\beta^2) + \mathcal{O}(\epsilon^3)
}{eq:s7a}
where we have already imposed the NLO consistency conditon from the main text, $\beta''+3\beta^\prime=\mathcal{O}(\beta)$. The last displayed term can again have either sign and thus lead to $f>1$ if $\epsilon$ is not small enough. 

Thus, subleading terms in $f$ put constraints on either the minimal dimension that one can consider for a given function $\beta$ to a given order in the expansion, or they can be used to constrain the function $\beta$ for a given value of the dimension.

\begin{figure}[!htb]
\includegraphics[height=0.21\linewidth]{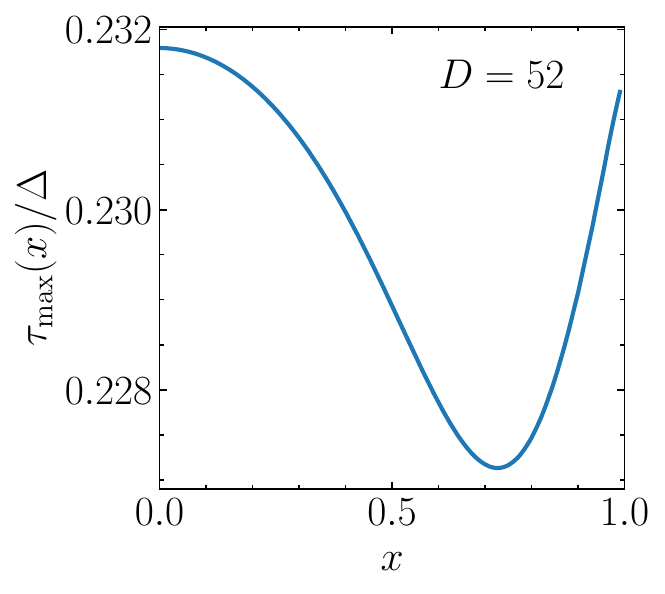}
\includegraphics[height=0.21\linewidth]{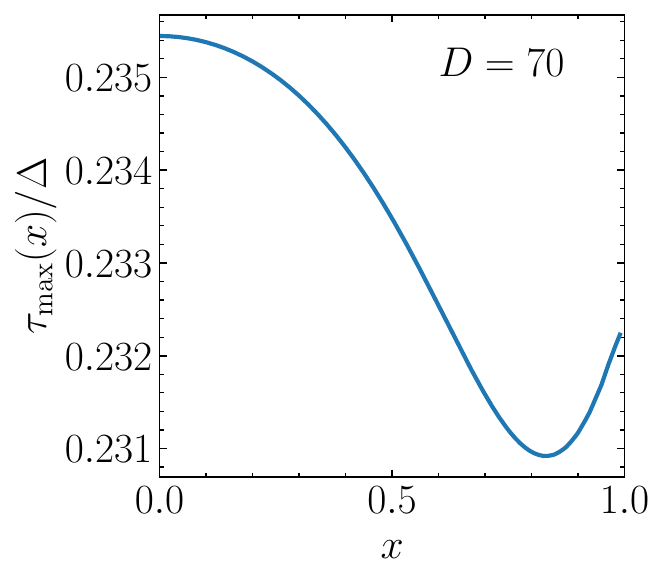}
\includegraphics[height=0.21\linewidth]{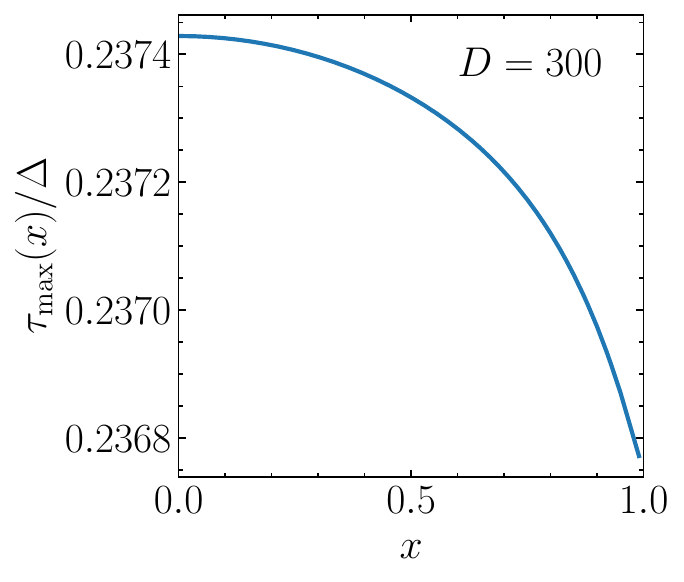}
\includegraphics[height=0.21\linewidth]{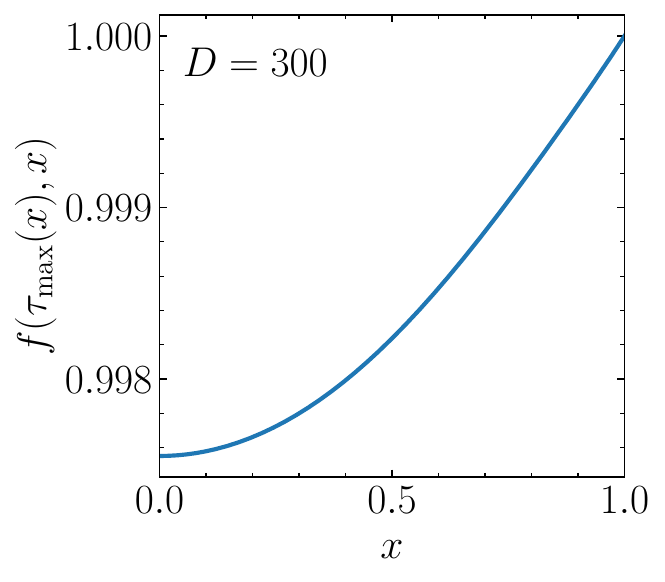}
\caption{Maxima of the SSH function $f$ at NNLO for different values of $D$ for the example \eqref{eq:example}.}
\label{fig:minmax}
\end{figure}

We conclude this section with plots of the locations of maxima of $f$ for the example in the main text to NNLO, see the left three plots in Fig.~\ref{fig:minmax}. The resulting curves match qualitatively with numerical data for $D\in(3.5,5.5)$. In particular, we recover the feature that for high dimensions the earliest time of the maximum-curve lies at or near the SSH, while at smaller values of the dimension it moves towards smaller values of $x$. The right plot in Fig.~\ref{fig:minmax} shows the decay of the maximum value of $f$ towards the center for the same example, in agreement with \eqref{eq:fmax}.

\end{document}